\documentclass[12pt]{article}
\usepackage{a4}
\usepackage{amssymb}
\usepackage{cite}
\usepackage{epsf}\usepackage{amssymb} \usepackage{cite}
\usepackage{amsthm}
\usepackage{amscd}
\usepackage{amsfonts}
\usepackage{amsmath}
\usepackage{hyperref}
\newcommand{\be}{\begin{equation}}
\newcommand{\ee}{\end{equation}}
\newcommand{\bea}{\begin{eqnarray}}
\newcommand{\eea}{\end{eqnarray}}

\renewcommand{\imath}{\mathrm{i}}

\begin{document}
\thispagestyle{empty}
\def\thefootnote{\fnsymbol{footnote}}
\begin{center}\Large
On relation of the genus one
Moore-Seiberg identity to the Baxter Q-operator in the hyperbolic Ruijsenaars model\\
\end{center}

\vskip 0.2cm

\begin{center}
Elena Apresyan$^{1}$\footnote{e.apresyan@aanl.am }
and
Gor Sarkissian$^{1,2}$
\footnote{sarkissn@theor.jinr.ru, gor.sargsyan@yerphi.am}
\end{center}
\begin{center}
$^1$Yerevan Physics Institute,\\
Alikhanian Br. 2, 0036\, Yerevan, Armenia
\end{center}
\begin{center}
and
\end{center}
\begin{center}
$^2$ Bogoliubov Laboratory of Theoretical Physics, JINR,\\
Dubna, Moscow region, 141980 Russia\\
\end{center}

\vskip 1.5em

\begin{abstract} \noindent
In this paper we show how the Baxter Q-operator and the product formula for eigenfunctions of two-particle hyperbolic Ruijsenaars system can be derived
from the genus one Moore-Seiberg duality identity in two-dimensional Liouville conformal field theory.
We expect that this relation would reveal genuine role of the Moore-Seiberg identity in integrable systems.
\end{abstract}

\newpage
\tableofcontents
\newpage

\section{Introduction}
The last years witnessed outburst of interest to Ruijsenaars wave function \cite{R11}.
This function appears in many cases in various branches of the theoretical and mathematical physics, e.g. in theory of integrable systems \cite{R11,Belousov:2023sat}, in two-dimensional conformal field theory \cite{Vartanov:2013ima}, in $4D$ and $3D$ supersymmetric gauge theories \cite{Hosomichi:2010vh,BKK} etc.  It is eigenfunction of the Hamiltonian of the two-particle relativistic Calogero-Sutherland
or Ruijsenaars model in the center of mass frame given by the finite-difference operator\footnote{In \cite{R11} the letter $b$ is used instead of $g$ here. But since the letter $b$ appears
in this paper in other role, we changed the notations to avoid a mess.}
 \cite{Rbanff,R11}:

\be\label{a55}
A_{\pm}(g;x)B(g;x,y)=2\cosh(\pi y/a_{\pm})B(g;x,y)\, ,
\ee

\be\label{a77}
A_{\pm}(g;x)={\sinh(\pi(x-\imath g)/a_{\pm})\over  \sinh(\pi x/a_{\pm})}T^{x}_{\imath a_{\mp}}+
{\sinh(\pi(x+\imath g)/a_{\pm})\over  \sinh(\pi x/a_{\pm})}T^{x}_{-\imath a_{\mp}}\, ,
\ee
where $T^z_c$ are translation operators:
$
(T^z_cf)(z)\equiv f(z-c),\; c\in \mathbb{C^*}\, ,
$
and
\be\label{rbg}
B(g;x,y)=G(a_{+},a_{-};\imath a-\imath g)^2\int_{\mathbb{R}}
dz G(a_{+},a_{-};\pm z\pm y/2-\imath a+\imath g/2)e^{\imath\alpha zx}\, .
\ee
Here we use notations $f(x\pm y)=f(x+y)f(x-y)$, $\alpha\equiv 2\pi/a_{+}a_{-}$, $a\equiv (a_{+}+a_{-})/2$, and  $G(a_{+},a_{-};y)$ is the hyperbolic gamma function given
by the integral representation:
\be\label{sbs1}
G(a_{+},a_{-};y)=\exp\left(\imath\int_0^{\infty}{dx\over x}\left({\sin 2yx\over 2\sinh(a_{+}x)
\sinh(a_{-}x)}-{y\over a_{+}a_{-}x}\right)\right)\, .
\ee
The function $B(g;x,y)$ can be considered as $q$-hypergeometric function \cite{NU,Ponsot:2000mt}.

It is shown in \cite{R11} that the function $B(g;x,y)$ possesses the nice self-duality property:
\be\label{dmual}
B(g;x,y)=G(a_{+},a_{-};\imath a-\imath g)^2B(2a-g;y,x)\, .
\ee
In applications to two-dimensional conformal field theory and three-dimensional $N=2$ supersymmetric gauge theories
appears  slightly different $S_b(x)$ hyperbolic gamma function reviewed in appendix A. Comparing \eqref{sbs1} and \eqref{sbs} in appendix
one sees easily that $G(a_{+},a_{-};y)$ and $S_b(x)$ related by the following equations:
\be
a_{+}=b,\quad a_{-}=b^{-1}, \quad G(b,b^{-1};y)=S_b\left({Q\over 2}-\imath y\right)\, ,
\ee
where
\be
Q=b+b^{-1}=2a\, .
\ee

Let us express the function \eqref{rbg} in terms of the function $S_b(x)$. Replacing $y$ by $\imath \lambda$, $x$ by $\imath x$,
changing the integration variable $z$ by $\imath z$,
also changing prefactor in front of integral from $G(a_{+},a_{-};\imath a-\imath g)^2=S_b(Q-g)^2$ to $S_b(Q-g)=1/S_b(g)$,
 and denoting derived in this way function by $F^g_{\lambda}(x)$,  we have:
\bea\label{fusik}
&&F^g_{\lambda}(x)={1\over S_b(g)}\int_{-\imath\infty}^{\imath\infty}S_b(\pm z\pm\lambda/2+g/2)
e^{2\imath\pi zx}  {dz\over \imath}\, .
\eea
Equations \eqref{a55}-\eqref{a77} in these variables take the form
\bea\label{ruseg2}
H\left[F^g_{\lambda}\right](x)=2\cos \pi b\lambda F^g_{\lambda}(x)\, ,
\eea
where
\bea\label{ruseg1}
H={\sin\pi b(x-g)\over \sin \pi x b}e^{-b\partial_x}  +
{\sin \pi b(x+g)\over \sin \pi b x}e^{b\partial_x}\, ,
\eea
and the second equation derived by the replacement $b\to 1/b$.

The self-duality property \eqref{dmual} in terms of the function $F^g_{\lambda}(x)$ reads:
\be\label{sdul}
F^g_{\lambda}(x)=F^{Q-g}_{x}(\lambda).
\ee
It is shown in \cite{halrus} that the function \eqref{fusik} enjoys the following product formula:
\bea\label{prodfo}
4F^g_{\lambda}(x_1)F^g_{\lambda}(x_2)=S_b(g)\int_{-\textup{i}\infty}^{\textup{i}\infty}
{S_b(g\pm z) \over S_b(\pm z)}
S_b((\bar{g}\pm z\pm x_1\pm x_2)/2)F^g_{\lambda}(z){dz\over \imath }\, ,
 \eea
 where $\bar{g}=Q-g$.
Multiplying both sides of \eqref{prodfo} by $e^{2\pi \imath\rho x_1}$ and integrating over $x_1$, one can rewrite formula \eqref{prodfo} in the form:
\bea\label{prodfo2}
&&{4S_b(\pm \rho\pm\lambda/2+g/2)\over\left(S_b(g)\right)^2} F^g_{\lambda}(x_2)\nonumber\\
&&=\int_{-\textup{i}\infty}^{\textup{i}\infty}\int_{-\textup{i}\infty}^{\textup{i}\infty}e^{2\pi \imath\rho x_1}
{S_b(g\pm z) \over S_b(\pm z)}
S_b((\bar{g}\pm z\pm x_1\pm x_2)/2)F^g_{\lambda}(z)dz dx_1.
 \eea
 This shows that if we define an integral operator
 \be\label{dbkk}
 [Q^{\rho}\phi](x)=\int_{-\textup{i}\infty}^{\textup{i}\infty}\int_{-\textup{i}\infty}^{\textup{i}\infty}e^{2\pi \imath\rho y_1}
{S_b(g\pm y_2) \over S_b(\pm y_2)}
S_b((\bar{g}\pm y_2\pm y_1\pm x)/2)\phi(y_2)dy_1 dy_2\, ,
  \ee
 the function \eqref{fusik} appears to be its eigen-vector:
 \be
  [Q^{\rho}F^g_{\lambda}](x)={4S_b(\pm \rho\pm\lambda/2+g/2)\over\left(S_b(g)\right)^2} F^g_{\lambda}(x)\, .
 \ee
 It is found in \cite{Belousov:2023qgn,Belousov:2023sat,halrus}, that $Q^{\rho}$ operators commute with the Hamiltonian \eqref{ruseg1}
 and with each other
 \be
 [Q^{\rho},H]=0 \quad {\rm and} \quad [Q^{\rho},Q^{\lambda}]=0\, ,
 \ee
and therefore can be considered as the Baxter $Q$-operators. It is shown in \cite{halrus,Belousov:2023sat}, that
in certain nonrelativistic limit the formula \eqref{prodfo} yields the product formula and the Baxter $Q$-operators
for eigenfunctions of the two-particle Calogero-Sutherland model. It was demonstrated in  \cite{Belousov:2025cms}, that in the complex limit $b\to \mathrm{i}$ eq. \eqref{prodfo} becomes product formula and the Baxter $Q$-operators for the complex hypergeometric functions \cite{Neretin:2019xze}.

 On the other hand the function \eqref{fusik} appears in two-dimensional conformal Liouville field theory as matrix
 of modular transformation $S_{\beta_1\beta_2}(\beta_3)$  of one-point conformal blocks on a torus \cite{Teschner:2003at}.
 It was established in \cite{Moore:1988uz,Moore:1988qv,Moore:1989vd} that the matrix of modular transformations of the one-point blocks
 should satisfy the genus one Moore-Seiberg (MS) identity:
\bea\label{msrel}
&&S_{\beta_1\beta_2}(\beta_3)\int_{\mathbb{S}} d\beta_4 F_{\beta_3,\beta_4}\left[\begin{array}{cc}
\beta_2&\alpha_1\\
\beta_2&\alpha_2 \end{array}\right]e^{2\pi \imath (\Delta_{\beta_4}-\Delta_{\beta_2})}
F_{\beta_4,\beta_5}\left[\begin{array}{cc}
\alpha_2&\alpha_1\\
\beta_2&\beta_2 \end{array}\right]\\ \nonumber
&&=\int_{\mathbb{S}} d\beta_6F_{\beta_3,\beta_6}\left[\begin{array}{cc}
\beta_1&\alpha_1\\
\beta_1&\alpha_2 \end{array}\right]e^{\pi \imath (\Delta_{\alpha_1}+\Delta_{\alpha_2}-\Delta_{\beta_5})}
F_{\beta_1,\beta_5}\left[\begin{array}{cc}
\alpha_1&\alpha_2\\
\beta_6&\beta_6 \end{array}\right]S_{\beta_6\beta_2}(\beta_5)\, ,
\eea
where $F_{\beta_3,\beta_4}\left[\begin{array}{cc}
\beta_2&\alpha_1\\
\beta_2&\alpha_2 \end{array}\right]$ is fusion matrix and $\Delta_{\alpha}$ is conformal weight of the primary labelled by $\alpha$.
In the case of $2D$ conformal Liouville field theory fusion matrix was computed in  \cite{Ponsot:1999uf,Ponsot:2000mt}:
\be\label{fff}
F_{\alpha_s,\alpha_t}\left[\begin{array}{cc}
\alpha_3&\alpha_2\\
\bar{\alpha_4} &\alpha_1 \end{array}\right]=
{S_b(\alpha_s+\alpha_2-\alpha_1)S_b(\alpha_1+\alpha_t-\alpha_4)\over
S_b(\alpha_t+\alpha_2-\alpha_3)S_b(\alpha_3+\alpha_s-\alpha_4)}
|S_b(2\alpha_t)|^2J_h(\beta_a^{\circ},\gamma_a^{\circ})\, ,
\ee
where as before $\bar{\alpha_4}=Q-\alpha_4$,  $Q=b+{1\over b}$ and
\be
\Delta_{\alpha}=\alpha(Q-\alpha)\quad {\rm and}\quad \mathbb{S}={Q\over 2}+\mathrm{i}\mathbb{R}.
\ee
Here $J_h(\beta_a^{\circ},\gamma_a^{\circ})$ is the  hyperbolic hypergeometric function, introduced in \cite{Rbanff,Ponsot:2000mt,Ponsot:1999uf}:
\be\label{jhptr}
J_h(\underline{\beta}^{\circ},\underline{\gamma}^{\circ})
=\int_{-\textup{i}\infty}^{\textup{i}\infty}\prod_{a=1}^4S_b(z+\gamma_a^{\circ})
S_b(-z+\beta_a^{\circ}) {dz\over \imath}\, .
\ee

The variables $\gamma_a^{\circ},\beta_a^{\circ}, a=1,2,3,4$,
in the Ponsot-Teschner parametrization are given by the formulae:
\be\label{ptparam}
\begin{aligned}
&\gamma_1^{\circ}=-Q/2+\alpha_3-\alpha_4\, ,\\
&\gamma_2^{\circ}=-Q/2+\alpha_1-\alpha_2\, ,\\
&\gamma_3^{\circ}=Q/2-\alpha_3-\alpha_4\, ,\\
&\gamma_4^{\circ}=Q/2-\alpha_1-\alpha_2\, ,
\end{aligned}\qquad
\begin{aligned}
&\beta_1^{\circ}=Q/2+\alpha_s\, ,\\
&\beta_2^{\circ}=Q/2-\alpha_t+\alpha_4+\alpha_2\, ,\\
&\beta_3^{\circ}=-Q/2+\alpha_t+\alpha_4+\alpha_2\, ,\\
&\beta_4^{\circ}=3Q/2-\alpha_s\, .
\end{aligned}
\ee
They satisfy the balancing condition $\sum_{a=1}^4(\gamma_a^{\circ}+\beta_a^{\circ})=2Q$.

Now we are going to explain how the equations \eqref{fusik}-\eqref{ruseg2} can be derived as particular cases of the relation \eqref{msrel}.

Setting in eq. \eqref{msrel} $\beta_1=\beta_3=0$  one obtains an
explicit expression of the $S$-matrix in terms of the fusion matrix and conformal weights  \cite{Moore:1989vd,Vartanov:2013ima}:
\bea\label{behr}
&&S_{0\beta_2}\int_{\mathbb{S}}d\beta_4F_{0,\beta_4}\left[\begin{array}{cc}
\beta_2&\alpha\\
\beta_2&\alpha \end{array}\right]e^{2\pi \imath \Delta_{\beta_4}}
F_{\beta_4,\beta_5}\left[\begin{array}{cc}
\alpha&\alpha\\
\beta_2&\beta_2 \end{array}\right]\\ \nonumber
&&=e^{\pi \imath (2\Delta_{\alpha}-\Delta_{\beta_5}+2\Delta_{\beta_2})}
F_{0,\beta_5}\left[\begin{array}{cc}
\alpha&\alpha\\
\alpha&\alpha \end{array}\right]S_{\alpha\beta_2}(\beta_5)\, .
\eea

Here we used that $S_{\alpha\beta_2}(\beta_5)$ in the limit $\beta_5\to 0$ becomes matrix $S_{\alpha\beta_2}$ of the modular transformation of characters and $S_{0\beta_2}$ in \eqref{behr} is the corresponding element of this matrix. The formula (\ref{behr}) was used in \cite{Vartanov:2013ima} to calculate $S_{\beta_6\beta_2}(\beta_5)$ in the Liouville field theory with the result
\be\label{sll}
S_{\alpha\beta_2}(\beta_5)=
S_{0\beta_2}e^{{\imath\pi\over 2}\Delta_{\beta_5}}F^{\beta_5}_{2\beta_2-Q}(2\alpha-Q)\, ,
\ee
where
\be
S_{0\beta_2}={1\over \sqrt{2}S_b(\pm 2(\beta_2-Q/2))}=-2\sqrt{2}\sin{\pi (2\beta_2-Q)\over b}\sin\pi (2\beta_2-Q)b\, .
\ee
So we see that the one-point matrix of modular transformations is given indeed essentially by function \eqref{fusik}.
If now we set in \eqref{msrel} $\alpha_1$ equal to degenerate value $\alpha_1=-b/2$  \cite{Nemkov:2015zha} and take into account
the fusion rule of a generic primary $V_{\alpha}$ with the degenerate primary $V_{-b/2}$:
\be
V_{\alpha}V_{-b/2}\sim V_{\alpha-b/2}+V_{\alpha+b/2}\, ,
\ee
we will obtain that $\beta_{3,4,5,6}$ can take only two values:
\be\nonumber
\beta_3=\alpha_2+s_1b/2\, ,\quad \beta_4=\beta_2+s_2b/2\, ,\quad \beta_5=\alpha_2+s_3b/2\, ,\quad \beta_6=\beta_1+s_4b/2\, ,
\ee
where $s_{1,2,3,4}=\pm 1$. Taking this into account one can write for \eqref{msrel} with one degenerate entry
\bea\nonumber
&&S_{\beta_1\beta_2}(\beta_3)\sum_{s_2=\pm} F_{\alpha_2+s_1b/2,\beta_2+s_2b/2}\left[\begin{array}{cc}
\beta_2&-b/2\\
\beta_2&\alpha_2 \end{array}\right]e^{-2\pi \imath (\Delta_{\beta_4}-\Delta_{\beta_2})}\\ \nonumber
&& \times F_{\beta_2+s_2b/2,\alpha_2+s_3b/2}\left[\begin{array}{cc}
\alpha_2&-b/2\\
\beta_2&\beta_2 \end{array}\right]\quad \quad\\ \nonumber
&&=\sum_{s_4=\pm} F_{\alpha_2+s_1b/2,\beta_1+s_4b/2}\left[\begin{array}{cc}
\beta_1&-b/2\\
\beta_1&\alpha_2 \end{array}\right]e^{-\pi \imath (\Delta_{-b/2}+\Delta_{\alpha_2}-\Delta_{\beta_5})}\\ \nonumber
&& \times F_{\beta_1,\alpha_2+s_3b/2}\left[\begin{array}{cc}
-b/2&\alpha_2\\
\beta_1+s_4b/2&\beta_1+s_4b/2 \end{array}\right]S_{\beta_1+s_4b/2\beta_2}(\beta_5)\, .
\eea
Inserting here the corresponding values of the fusion matrix elements with one degenerate entry, which are explicitly calculable, see for example
\cite{Belavin:1984vu,Ponsot:1999uf,Nemkov:2015zha}, we obtain
\be
H(S_{\alpha\beta_2}(\beta_5))=2\cos\pi b(Q-2\beta_2)S_{\alpha\beta_2}(\beta_5)\, ,
\ee
where $H$ is the finite-difference operator:
\bea
H={\sin\pi b(2\alpha-Q-\beta_5)\over \sin \pi b(2\alpha-Q)}e^{-{b\over 2}\partial_{\alpha}}+
{\sin\pi b(2\alpha-Q+\beta_5)\over \sin \pi b (2\alpha-Q)}e^{{b\over 2}\partial_{\alpha}}\, .
\eea

Thus we have shown that equations \eqref{ruseg2} and \eqref{ruseg1} are particular cases of \eqref{msrel}.

Now we will show that the formula \eqref{prodfo} is consequence of \eqref{msrel} as well.
In the rest of paper we show that the Moore-Seiberg identity \eqref{msrel} reduces to the  product formula \eqref{prodfo} upon setting
\be\label{mkultra}
\alpha_2-\alpha_1=\beta_5=\beta_3\, .
\ee
In fact we find that both sides of \eqref{msrel} under this condition contain the same divergent term $S_b(\beta_3+\alpha_1-\alpha_2)$,
which therefore peacefully drops. So, strictly speaking, we set $\beta_5=\beta_3$ and calculate the limit $\varepsilon=\beta_3+\alpha_1-\alpha_2=\beta_5+\alpha_1-\alpha_2\to 0$.

\section{Evaluation of the RHS of the MS identity}
Now we start evaluation of the Moore-Seiberg identity \eqref{msrel} under the condition \eqref{mkultra}. In
this section we calculate the right-hand side of \eqref{msrel}, and in the next section we calculate the left-hand side of it and equate
with the computed here  right-hand side.

Let us at the beginning compute the first fusion matrix element appearing in the right-hand side of eq.\eqref{msrel}:
\be\label{perva}
F_{\beta_3,\beta_6}\left[\begin{array}{cc}
\beta_1&\alpha_1\\
\beta_1&\alpha_2 \end{array}\right]\, .
\ee
In notations \eqref{fff} this means
\bea\label{nnn1}
\alpha_1\equiv \alpha_2 ,\quad \alpha_2\equiv \alpha_1,\quad \alpha_3\equiv \beta_1,\quad \alpha_4\equiv Q-\beta_1, \quad
\alpha_s\equiv \beta_3,\quad \alpha_t\equiv \beta_6\, .
\eea
Inserting \eqref{nnn1} in \eqref{fff} we compute the prefactor for the corresponding element
\be\label{prefik1}
{S_b(\varepsilon)S_b(\alpha_2+\beta_6-Q+\beta_1)\over
S_b(\beta_6+\alpha_1-\beta_1)S_b(2\beta_1+\beta_3-Q)}
|S_b(2\beta_6)|^2\, ,
\ee
and putting \eqref{nnn1} in \eqref{ptparam} we get the corresponding Ponsot-Teschner parameters:
\be\label{ptparam1}
\begin{aligned}
&\gamma_1^{\circ}=-3Q/2+2\beta_1\, ,\\
&\gamma_2^{\circ}=-Q/2+\alpha_2-\alpha_1\, ,\\
&\gamma_3^{\circ}=-Q/2\, ,\\
&\gamma_4^{\circ}=Q/2-\alpha_2-\alpha_1\, ,
\end{aligned}\qquad
\begin{aligned}
&\beta_1^{\circ}=Q/2+\beta_3\, ,\\
&\beta_2^{\circ}=3Q/2-\beta_6-\beta_1+\alpha_1\, ,\\
&\beta_3^{\circ}=Q/2+\beta_6-\beta_1+\alpha_1\, ,\\
&\beta_4^{\circ}=3Q/2-\beta_3\, .
\end{aligned}
\ee

The condition \eqref{mkultra}
implies $\gamma_2^{\circ}+\beta_4^{\circ}=Q$ and using the reflection formula \eqref{refl} and the formula \eqref{sinteg} in appendix A
we can explicitly calculate the integral $J^{I}_h(\underline{\beta}^{\circ},\underline{\gamma}^{\circ})$ entering in the mentioned fusion matrix element:
\bea\label{inty1}
&&J^{I}_h(\underline{\beta}^{\circ},\underline{\gamma}^{\circ})=S_b(-Q+2\beta_1+\beta_3)S_b(-\beta_6+\beta_1+\alpha_1)
S_b(-Q+\beta_6+\beta_1+\alpha_1)S_b(\beta_3)
\nonumber\\
&&\times S_b(Q-\beta_6-\beta_1+\alpha_1)S_b(\beta_6-\beta_1+\alpha_1)S_b(Q-2\alpha_1)S_b(2Q-\beta_6-\beta_1-\alpha_2)\nonumber\\
&&\times S_b(Q+\beta_6-\beta_1-\alpha_2)\, .
\eea
Multiplying \eqref{prefik1} and \eqref{inty1} we receive after some cancellations:
\begin{eqnarray}\label{f1}
&&F_{\beta_3,\beta_6}\left[\begin{array}{cc}
\beta_1&\alpha_1\\
\beta_1&\alpha_2 \end{array}\right]=
S_b(\varepsilon)
|S_b(2\beta_6)|^2
 S_b(-\beta_6+\beta_1+\alpha_1)S_b(-Q+\beta_6+\beta_1+\alpha_1)\nonumber \\
&& \times S_b(\beta_3)S_b(Q-\beta_6-\beta_1+\alpha_1)S_b(Q-2\alpha_1)S_b(Q+\beta_6-\beta_1-\alpha_2)\, .\qquad\qquad
\end{eqnarray}
Let us turn to the second necessary element of the fusion matrix on the right-hand side of \eqref{msrel}:
$
F_{\beta_1,\beta_5}\left[\begin{array}{cc}
\alpha_1&\alpha_2\\
\beta_6&\beta_6 \end{array}\right]\, .
$
In notations \eqref{fff} this means
\bea
\alpha_1\equiv \beta_6 ,\quad \alpha_2\equiv \alpha_2,\quad \alpha_3\equiv \alpha_1,\quad \alpha_4\equiv Q-\beta_6,\quad
\alpha_s\equiv \beta_1,\quad \alpha_t\equiv \beta_5\, .
\eea
In this case for the prefactor in \eqref{fff} we obtain:
\be\label{prefik2}
{S_b(\beta_1+\alpha_2-\beta_6)S_b(2\beta_6+\beta_5-Q)\over
S_b(2\beta_5)S_b(\alpha_1+\beta_1-Q+\beta_6)}
|S_b(2\beta_5)|^2\, ,
\ee
and for the parameters \eqref{ptparam} we get
\be\label{ptparam2}
\begin{aligned}
&\gamma_1^{\circ}=-3Q/2+\alpha_1+\beta_6\, ,\\
&\gamma_2^{\circ}=-Q/2+\beta_6-\alpha_2\, ,\\
&\gamma_3^{\circ}=-Q/2-\alpha_1+\beta_6\, ,\\
&\gamma_4^{\circ}=Q/2-\beta_6-\alpha_2\, ,
\end{aligned}\qquad
\begin{aligned}
&\beta_1^{\circ}=Q/2+\beta_1\, ,\\
&\beta_2^{\circ}=3Q/2-\beta_5-\beta_6+\alpha_2\, ,\\
&\beta_3^{\circ}=Q/2+\beta_5-\beta_6+\alpha_2\, ,\\
&\beta_4^{\circ}=3Q/2-\beta_1\, .
\end{aligned}
\ee
The condition \eqref{mkultra}
implies $\gamma_3^{\circ}+\beta_2^{\circ}=Q$ and, as before, using formulas \eqref{refl} and \eqref{sinteg} in appendix, we obtain for the integral in \eqref{fff}:
\bea\label{inty2}
&&J^{II}_h(\underline{\beta}^{\circ},\underline{\gamma}^{\circ})=S_b(-Q+\alpha_1+\beta_1+\beta_6)S_b(-Q+2\alpha_2)S_b(\alpha_1+\beta_6-\beta_1)
S_b(\beta_6-\alpha_2+\beta_1)S_b(\beta_5)\nonumber\\
&&\times S_b(Q+\beta_6-\alpha_2-\beta_1)
S_b(Q-\beta_6-\alpha_2+\beta_1)S_b(Q+\beta_5-2\beta_6)S_b(2Q-\beta_6-\alpha_2-\beta_1)\, .
\eea

Collecting \eqref{prefik2} and \eqref{inty2} after some cancellations we obtain the second necessary fusion matrix element:
\begin{eqnarray}\label{f2}
&&F_{\beta_1,\beta_5}\left[\begin{array}{cc}
\alpha_1&\alpha_2\\
\beta_6&\beta_6 \end{array}\right]=
{S_b(2\beta_6+\beta_5-Q)\over
S_b(2\beta_5)}
|S_b(2\beta_5)|^2
S_b(-Q+2\alpha_2)S_b(\alpha_1+\beta_6-\beta_1)\\
&&\times S_b(\beta_6-\alpha_2+\beta_1)S_b(\beta_5)
S_b(Q-\beta_6-\alpha_2+\beta_1)S_b(Q+\beta_5-2\beta_6)S_b(2Q-\beta_6-\alpha_2-\beta_1)\, .\nonumber
\end{eqnarray}

Taking into account \eqref{f1} and \eqref{f2}, we derive for the right-hand side of \eqref{msrel}:

\begin{eqnarray}\label{rhs}
&&\int_{\mathbb{S}} d\beta_6 F_{\beta_3,\beta_6}\left[\begin{array}{cc}
\beta_1&\alpha_1\\
\beta_1&\alpha_2 \end{array}\right]e^{\pi \imath (\Delta_{\alpha_1}+\Delta_{\alpha_2}-\Delta_{\beta_5})}
F_{\beta_1,\beta_5}\left[\begin{array}{cc}
\alpha_1&\alpha_2\\
\beta_6&\beta_6 \end{array}\right]S_{\beta_6\beta_2}(\beta_5)\nonumber\\
&&={\mathcal I}_1\times {\mathcal A}\times e^{\pi \imath (\Delta_{\alpha_1}+\Delta_{\alpha_2}-\Delta_{\beta_5})}\, ,
\end{eqnarray}
where
\bea\label{mtha}
{\mathcal A}=|S_b(2\beta_5)|^2S_b(\varepsilon)(S_b(\beta_3))^2S_b(Q-2\alpha_1)
S_b(Q-2\beta_5)S_b(-Q+2\alpha_2)\, ,\quad
\eea
and
\begin{eqnarray}
\label{belint}
&& {\mathcal I}_1=\int_{\mathbb{S}} d\beta_6 |S_b(2\beta_6)|^2S_b(2\beta_6+\beta_5-Q)S_b(Q+\beta_5-2\beta_6)\nonumber \\
&&\times
S_b(Q+\beta_6-\alpha_2-\beta_1)S_b(-Q+\beta_6+\alpha_1+\beta_1)S_b(\beta_6-\beta_1+\alpha_1)S_b(\beta_6-\alpha_2+\beta_1)\nonumber\\
&&\times S_b(-\beta_6+\beta_1+\alpha_1)S_b(Q-\beta_6+\beta_1-\alpha_2)S_b(Q-\beta_6-\beta_1+\alpha_1)\nonumber\\
&&\times S_b(2Q-\beta_6-\beta_1-\alpha_2)
S_{\beta_6\beta_2}(\beta_5)\, .
\end{eqnarray}
Let us introduce new variables $z$, $y$, $t$, $u$:
\be\label{newvar}
\beta_6=z+{Q\over 2}\, ,\quad \beta_1={Q\over 2}-y\, ,\quad \alpha_1={Q\over 2}-t-{\beta_5\over 2}\, ,\quad \beta_2={Q\over 2}-u\, ,
\ee
The form of the spectrum $\mathbb{S}={Q\over 2}+\mathrm{i}\mathbb{R}$ requires $z$ to be pure imaginary: $z\in \mathrm{i}\mathbb{R}$.
Condition \eqref{mkultra} implies that in these variables
\be\label{newvar3}
\alpha_2={Q\over 2}-t+{\beta_5\over 2}\, .
\ee

Now recalling \eqref{sll}, it is straightforward to check that the integral \eqref{belint} in these variables takes the compact form:
\be\label{int8}
{\mathcal I}_1=S_{0\beta_2}e^{{\imath\pi\over 2}\Delta_{\beta_5}}\int_{-\imath\infty}^{\imath\infty} {S_b(\beta_5\pm 2z) \over S_b(\pm 2z)} S_b(\bar{\beta}_5/2\pm z\pm y\pm t)F^{\beta_5}_{2u}(2z)
dz\, ,
\ee
where $\bar{\beta}_5=Q-\beta_5$.

\section{Evaluation of the LHS of the MS identity}

Turn to the first  element of the fusion matrix on the left-hand side of \eqref{msrel}:
$
F_{\beta_3,\beta_4}\left[\begin{array}{cc}
\beta_2&\alpha_1\\
\beta_2&\alpha_2 \end{array}\right]\, .
$
It is easy to see that this element of the fusion matrix can be derived from the first matrix element \eqref{perva}, computed in the previous section, by the replacement
$\beta_6\to \beta_4$ and $\beta_1\to \beta_2$. Therefore the final expression for this element can be obtained from \eqref{f1} by the same replacement:
\begin{eqnarray}\label{f3}
&&F_{\beta_3,\beta_4}\left[\begin{array}{cc}
\beta_2&\alpha_1\\
\beta_2&\alpha_2 \end{array}\right]=
S_b(\varepsilon)
|S_b(2\beta_4)|^2
S_b(-\beta_4+\beta_2+\alpha_1)S_b(-Q+\beta_4+\beta_2+\alpha_1)\nonumber\\
&&\times S_b(\beta_3)S_b(Q-\beta_4-\beta_2+\alpha_1)
S_b(Q-2\alpha_1)S_b(Q+\beta_4-\beta_2-\alpha_2)\, .\qquad
\end{eqnarray}

It remains to compute the last necessary element of the fusion matrix
$
F_{\beta_4,\beta_5}\left[\begin{array}{cc}
\alpha_2&\alpha_1\\
\beta_2&\beta_2 \end{array}\right]\, .
$
In notations of eq. \eqref{fff} this means
\bea\label{pupupu}
\alpha_1\equiv \beta_2 ,\quad \alpha_2\equiv \alpha_1,\quad \alpha_3\equiv \alpha_2,\quad \alpha_4\equiv Q-\beta_2,\quad
\alpha_s\equiv \beta_4,\quad \alpha_t\equiv \beta_5.
\eea
Using \eqref{pupupu} in this case for the prefactor in \eqref{fff} we derive
\be\label{prefik4}
{S_b(\beta_4+\alpha_1-\beta_2)S_b(2\beta_2+\beta_5-Q)\over
S_b(\varepsilon)S_b(\alpha_2+\beta_4-Q+\beta_2)}
|S_b(2\beta_5)|^2\, ,
\ee
and for the parameters \eqref{ptparam} we obtain

\be\label{ptparam4}
\begin{aligned}
&\gamma_1^{\circ}=-3Q/2+\alpha_2+\beta_2\, ,\\
&\gamma_2^{\circ}=-Q/2+\beta_2-\alpha_1\, ,\\
&\gamma_3^{\circ}=-Q/2-\alpha_2+\beta_2\, ,\\
&\gamma_4^{\circ}=Q/2-\beta_2-\alpha_1\, ,
\end{aligned}\qquad
\begin{aligned}
&\beta_1^{\circ}=Q/2+\beta_4\, ,\\
&\beta_2^{\circ}=3Q/2-\beta_5-\beta_2+\alpha_1\, ,\\
&\beta_3^{\circ}=Q/2+\beta_5-\beta_2+\alpha_1\, ,\\
&\beta_4^{\circ}=3Q/2-\beta_4\, .
\end{aligned}
\ee

Now recall the formula \eqref{ide1b} in appendix A.

Take there
\be
\nu_1=\gamma_1^{\circ}\, ,\quad \nu_2=\gamma_4^{\circ}\, ,\quad \nu_3=\gamma_2^{\circ}\, ,\quad \nu_4=\gamma_3^{\circ}\, ,
\ee

\be
\mu_1=\beta_1^{\circ}\, , \quad \mu_2=\beta_4^{\circ}\, , \quad \mu_3=\beta_2^{\circ}\, , \quad \mu_4=\beta_3^{\circ}\, .
\ee

For this choice $\eta$ in \eqref{ball8} takes the value $\eta=\alpha_1-\alpha_2$.

One can easily obtain that new parameters in argument of $J_h$ in the right-hand side of \eqref{ide1b} can be derived from \eqref{ptparam4}
by exchanging $\alpha_1\leftrightarrows \alpha_2$. After this exchange one can see that new parameters can be derived from
\eqref{ptparam2} by the replacement $\beta_6\to \beta_2$ and $\beta_1\to \beta_4$, therefore the integral can be obtained from
\eqref{inty2} by the replacement $\beta_6\to \beta_2$ and $\beta_1\to \beta_4$.
Taking the factors in front of integral in \eqref{ide1b} and the value of the integral derived by the mentioned replacement, after some cancellations we obtain:
\bea\label{inty4}
&&J^{IV}_h(\gamma_a^{\circ},\beta_a^{\circ})=
S_b(Q-2\beta_5)S_b(\varepsilon)S_b(-Q+2\alpha_2)S_b(\beta_2-\alpha_2+\beta_4)\nonumber \\
&&\times S_b(\beta_5)
S_b(Q+\beta_2-\alpha_2-\beta_4)
S_b(Q+\beta_5-2\beta_2)\, .
\eea
Finally putting together \eqref{prefik4} and \eqref{inty4} we derive the second element of the fusion matrix in the left-hand side of \eqref{msrel}:
\begin{eqnarray}\label{f4}
&&F_{\beta_4,\beta_5}\left[\begin{array}{cc}
\alpha_2&\alpha_1\\
\beta_2&\beta_2 \end{array}\right]=
{S_b(\beta_4+\alpha_1-\beta_2)S_b(2\beta_2+\beta_5-Q)\over
S_b(\alpha_2+\beta_4-Q+\beta_2)}
|S_b(2\beta_5)|^2
S_b(Q-2\beta_5)\nonumber \\
&&\times S_b(-Q+2\alpha_2)S_b(\beta_2-\alpha_2+\beta_4)S_b(\beta_5)
S_b(Q+\beta_2-\alpha_2-\beta_4)
S_b(Q+\beta_5-2\beta_2)\, .\qquad
\end{eqnarray}
Note that the divergent factor $S_b(\varepsilon)$ in denominator of  \eqref{prefik4} cancels with the same factor in \eqref{inty4}.
Inserting \eqref{f3} and \eqref{f4} in the left-hand side of \eqref{msrel} we obtain:
\bea\label{lhs}
&&S_{\beta_1\beta_2}(\beta_5)\int_{\mathbb{S}} d\beta_4F_{\beta_3,\beta_4}\left[\begin{array}{cc}
\beta_2&\alpha_1\\
\beta_2&\alpha_2 \end{array}\right]e^{2\pi \imath (\Delta_{\beta_4}-\Delta_{\beta_2})}F_{\beta_4,\beta_5}\left[\begin{array}{cc}
\alpha_2&\alpha_1\\
\beta_2&\beta_2 \end{array}\right]\nonumber\\
&&=S_{\beta_1\beta_2}(\beta_5)\times {\mathcal A}\times {\mathcal I}_2\times e^{-2\pi \imath \Delta_{\beta_2}}\, ,
\eea
where ${\mathcal A}$ is defined by \eqref{mtha}, and ${\mathcal I}_2$ is the integral
\bea\label{mmm}
&&{\mathcal I}_2=S_b(2\beta_2+\beta_5-Q)
S_b(Q+\beta_5-2\beta_2)\int_{\mathbb{S}} d\beta_4 e^{2\pi \imath \Delta_{\beta_4}}|S_b(2\beta_4)|^2S_b(-\beta_4+\beta_2+\alpha_1)\nonumber\\
&&\times S_b(Q-\beta_4-\beta_2+\alpha_1)
S_b(\beta_4+\alpha_1-\beta_2)S_b(\alpha_1+\beta_4-Q+\beta_2)
S_b(2Q-\alpha_2-\beta_4-\beta_2)\nonumber\\
&&\times S_b(Q+\beta_2-\alpha_2-\beta_4)S_b(\beta_2-\alpha_2+\beta_4)S_b(Q+\beta_4-\beta_2-\alpha_2)\, .
\eea
Let us closer look at the integral \eqref{mmm}. For this purpose introduce the variable $x$
\be
x=-{Q\over 2}+\beta_4\, ,\quad x\in \mathrm{i}\mathbb{R}\, ,
\ee
and remember the expression for $\beta_2$ via the variable $u$ given in \eqref{newvar}.
In these variables expression \eqref{mmm} can be compactly written as
\be
{\mathcal I}_2=e^{\imath Q^2\pi\over 2}S_b(\beta_5\pm 2u)
\int_{-\textup{i}\infty}^{\textup{i}\infty} {S_b(\alpha_1\pm u\pm x)S_b(Q-\alpha_2\pm u\pm x)
\over S_b(\pm 2x)}e^{-2\imath\pi x^2}dx\, .\\
\ee
It is easy to see that this expression up to factor $e^{\imath Q^2\pi\over 2}$ coincides with the right-hand side of formula \eqref{sss} in appendix.
Therefore using this formula we can equate  ${\mathcal I}_2$ with the left-hand side of \eqref{sss}:
\be\label{intros}
{\mathcal I}_2=2e^{\imath Q^2\pi\over 2}e^{2\imath\pi(\alpha_1(Q-\alpha_2)-u^2)}
\int_{-\imath\infty}^{\imath\infty}S_b(\alpha_1\pm p)
S_b(Q-\alpha_2\pm p)e^{4\imath\pi up}  dp\, .
\ee
Remembering the expressions \eqref{newvar} and \eqref{newvar3} for $\alpha_1$ and $\alpha_2$ via the variable $t$, the integral on the right-hand side of \eqref{intros}
takes the form
\bea
&&\int_{-\imath\infty}^{\imath\infty}S_b(\alpha_1\pm p)
S_b(Q-\alpha_2\pm p)e^{4\imath \pi up}  dp\nonumber \\
&&=\int_{-\imath\infty}^{\imath\infty}S_b\left(-t+{Q-\beta_5\over 2}\pm p\right)
S_b\left(t+{Q-\beta_5\over 2}\pm p\right)e^{4\imath\pi up}  dp\, .
\eea
Comparing with the definition of the function $F^g_{\lambda}(x)$ in eq. \eqref{fusik} we see
\be\label{intros2}
\int_{-\imath\infty}^{\imath\infty}S_b\left({Q-\beta_5\over 2}\pm t\pm p\right)
e^{4\imath\pi up}  dp=S_b(Q-\beta_5)F^{Q-\beta_5}_{2t}(2u)=S_b(Q-\beta_5)F^{\beta_5}_{2u}(2t),
\ee
where at the last step we used the self-duality property \eqref{sdul}.
Now we are ready to equate the left- \eqref{lhs} and right-hand sides \eqref{rhs} of \eqref{msrel}.
Cancelling the common (divergent) factor ${\mathcal A}$ and using \eqref{intros}-\eqref{intros2} we arrive at
\begin{eqnarray*}
2S_{\beta_1\beta_2}(\beta_5)S_b(Q-\beta_5)F^{\beta_5}_{2u}(2t)e^{\imath Q^2\pi\over 2}e^{-2\pi \imath \Delta_{\beta_2}}
e^{2\imath\pi(\alpha_1(Q-\alpha_2)-u^2)}
={\mathcal I}_1e^{\pi \imath (\Delta_{\alpha_1}+\Delta_{\alpha_2}-\Delta_{\beta_5})}\, .
\end{eqnarray*}

One can check that all the exponents get cancelled. Finally recalling \eqref{sll} and \eqref{int8} we derive

\be
2F^{\beta_5}_{2u}(2y)F^{\beta_5}_{2u}(2t)
=S_b(\beta_5)\int_{-\imath\infty}^{\imath\infty} {S_b(\beta_5\pm 2z) \over S_b(\pm 2z)} S_b(\bar{\beta}_5/2\pm z\pm y\pm t)F^{\beta_5}_{2u}(2z)
dz\, .
\ee
To obtain \eqref{prodfo} it remains to make obvious replacement $y\to {y\over 2}$, $t\to {t\over 2}$, $z\to {z\over 2}$, $u\to {u\over 2}$.

\section{Conclusion}
In this paper we have shown that the product formula of the two-particle Ruijsenaars wave functions, which coincides in fact with the
equation describing action of the Baxter Q-operators on the wave functions, can be derived as particular case of the genus one
Moore-Seiberg identity in the Liouville field theory.
Since  general N-particle hyperbolic Ruijsenaars system is related to the two-dimensional conformal Toda field theory \cite{Hosomichi:2010vh,BKK}
we can expect that similar relation should exist also between
Moore-Seiberg identity \eqref{msrel} in the conformal Toda field theory and N-particle hyperbolic Ruijsenaars system.
Another direction is the supersymmetric generalization of the hyperbolic Ruijsenaars system.
In our previous paper  \cite{Apresyan:2023vxl}, an attempt to construct supersymmetric generalization was made, where
we suggested a model whose eigen-functions are given by elements of one-point matrix of the modular transformation in $N=1$ super
Liouville conformal field theory. We think that similar studies of the  identity \eqref{msrel} in
$N=1$ super Liouville conformal field theory will bring to the corresponding product formula for the elements of one-point matrix of the modular transformation in $N=1$ super
Liouville conformal field theory.
We hope that deep understanding of the relation between the Moore-Seiberg identity and the Q-Baxter operators
can reveal genuine role of the  Moore-Seiberg identity in integrable systems.
\newpage

{\bf Acknowledgements.}
The work of Elena Apresyan was supported by  Armenian SCS grants 21AG‐1C024 and 24FP-1F039. We would like also to thank
N. M. Belousov, S. E. Derkachov and V. P. Spiridonov for many useful discussions.

\appendix

\section{Integral identities for hyperbolic gamma function}
The function $S_b(y)$ has the integral representation (see for review e.g. \cite{Spiridonov:2010em}):
\be\label{sbs}
S_b(y)=\exp\left(-\int_0^{\infty}{dx\over x}\left({\sinh(2y-b-b^{-1})x\over 2\sinh(bx)
\sinh(b^{-1}x)}-{2y-b-b^{-1}\over 2x}\right)\right)\, .
\ee
It enjoys the reflection property
\be\label{refl}
S_b(x)S_b(Q-x)=1\, ,
\ee
and obeys the shift equations:
\be\label{hp}
{S_b(y+b)\over S_b(y)}=2\sin\pi by\,  ,\quad
{S_b(y+b^{-1})\over S_b(y)}=2\sin{\pi y\over b}.
\ee
The poles and zeros of the function $S_b(y)$ are given by the relations :
\be\nonumber
y_{\rm poles}\in \{ -n_1b -n_2b^{-1}\},\; y_{\rm zeros}\in \{ (n_1+1)b +(n_2+1)b^{-1}\},
\ee
where
$ n_1,n_2 \in {\mathbb Z}_{\geq0}$.

The function $S_b(y)$ satisfies the following identity proved in \cite{Kashaev} (see also \cite{Belousov:2024cwe}):

\be\label{sinteg}
\int_{-\textup{i}\infty}^{\textup{i}\infty} {dx\over i}\prod_{i=1}^3 S_{b}(x+a_i)
S_{b}(-x+b_i)=\prod_{i,j=1} S_{b}(a_i+b_j)\, ,
\ee
where
\be
\sum_i(a_i+b_i)=Q\, .
\ee
The hyperbolic hypergeometric function \eqref{jhptr} enjoys the symmetry property \cite{BRS,Apresyan:2022erh}
\bea\label{ide1b}
&&
J_h(\underline{\mu},\underline{\nu})=\prod_{ j, k =1}^2S_b(\mu_j+\nu_k)\prod_{ j, k =3}^4S_b(\mu_j+\nu_k)
\\ \nonumber && \makebox[2em]{}
\times J_h(\mu_1,\mu_2,\mu_3-\eta,\mu_4-\eta,\nu_1+\eta,\nu_2+\eta,\nu_3,\nu_4),
\eea
where
\be\label{jhh7}
J_h(\underline{\mu},\underline{\nu})=\int_{-\textup{i}\infty}^{\textup{i}\infty}\prod_{a=1}^4S_b(\mu_a- z)
S_b(\nu_a+ z)\frac{dz}{\textup{i}}
\ee
with the parameters $\eta$, $\mu_a$, $\nu_a$ satisfying the conditions
\be\label{ball8}
\sum_{a=1}^4(\nu_a+\mu_a)=2Q, \quad {\rm and} \quad \eta=Q-\mu_1-\mu_2-\nu_1-\nu_2.
\ee

At the last part we need the integral identity \cite{bult,Apresyan:2023vxl}

\bea\label{sss}
&&2e^{2\imath\pi\left(\mu_1\mu_2-\alpha^2\right)}\int_{-\imath\infty}^{\imath\infty}\prod_{i=1,2}S_b(\mu_i\pm z)
e^{4\imath\pi z\alpha}dz
\\ \nonumber
&&= S_b(Q-\mu_1-\mu_2\pm 2\alpha)\int_{-\imath\infty}^{\imath\infty}{\prod_{i=1,2}S_b(\mu_i\pm\alpha\pm z)\over
S_b(\pm 2z)}e^{-2\imath\pi z^2}dz\, .
\eea

\end{document}